\documentclass[twocolumn,english,prl,floatfix,citeautoscript,nofootinbib]{revtex4}
\usepackage{amsbsy}
\usepackage{latexsym,epsfig,graphicx}
\usepackage{dcolumn}
\usepackage{subfigure}
\usepackage{comment}
\usepackage{color}
\usepackage[colorlinks,urlcolor=blue,citecolor=blue]{hyperref}
\usepackage{amstext}
\usepackage{amssymb}
\usepackage{setspace}

\begin{document}

\title{Self-adapted Floquet Dynamics of Ultracold Bosons in a Cavity}
\author{Xi-Wang Luo}
\author{Chuanwei Zhang}
\email{chuanwei.zhang@utdallas.edu}
\affiliation{Department of Physics, The University of Texas at Dallas, Richardson, Texas
75080-3021, USA}

\begin{abstract}
Floquet dynamics of a quantum system subject to periodic modulations of
system parameters provide a powerful tool for engineering new quantum matter
with exotic properties. While system dynamics are significantly altered, the
periodic modulation itself is usually induced externally and independent of
Floquet dynamics. Here we propose a new type of Floquet physics for a
Bose-Einstein condensate (BEC) subject to a shaken lattice generated inside
a cavity, where the shaken lattice and atomic Floquet bands are mutually
dependent, resulting in self-adapted Floquet dynamics. In particular, the
shaken lattice induces Floquet quasi-energy bands for the BEC, whose back
action leads to a self-adapted dynamical normal-superradiant phase
transition for the shaken lattice. Such self-adapted Floquet dynamics show
two surprising and unique features: \textit{i}) the normal-superradiant
phase transition possesses a hysteresis even without atom interactions;
\textit{ii}) the dynamical atom-cavity steady state could exist at free
energy maxima. The atom interactions strongly affect the phase transition of
the BEC from zero to finite momenta. Our results provide a powerful platform
for exploring self-adapted Floquet physics, which may open an avenue for
engineering novel quantum materials.
\end{abstract}

\maketitle

\emph{Introduction.---}Floquet physics has been extensively studied in solid
state, ultracold atomic, and photonic systems in recent years with
significant theoretical and experimental progress~\cite{lindner2011floquet,
wang2013observation, de2016monitoring, cayssol2013floquet,
hubener2017creating, PhysRevX.4.031027, holthaus2015floquet,
eckardt2017colloquium, kitagawa2012observation, fang2012realizing,
rechtsman2013photonic, fang2012photonic, li2014photonic, basov2017towards}.
In particular, ultracold atoms in periodically driven optical lattices
provide a highly controllable and disorder-free platform for studying
Floquet physics, yielding many interesting and important phenomena such as
coherent ac-induced tunneling and band coupling~\cite{PhysRevB.34.3625,
PhysRevLett.100.043602, PhysRevLett.100.040404, PhysRevLett.104.200403,
alberti2009engineering, PhysRevLett.95.170404, bakr2011orbital,
parker2013direct, PhysRevLett.114.055301, PhysRevLett.116.120403,
khamehchi2016spin}, the realization of gauge fields and topological bands~%
\cite{PhysRevLett.94.086803, PhysRevB.82.235114, PhysRevLett.111.185301,
PhysRevLett.107.255301, PhysRevLett.108.225304, struck2013engineering,
PhysRevLett.111.185302, atala2014observation, kennedy2015observation,
tai2017microscopy, jotzu2014experimental, aidelsburger2015measuring}, and
the dynamical control of expansion and quantum phase transition of bosonic
systems~\cite{PhysRevLett.95.260404, PhysRevLett.99.220403,
PhysRevLett.102.100403, struck2011quantum}, etc. These previous studies on
Floquet physics assumed that system parameter modulations (e.g., the shaking
or moving optical lattices) are determined solely by external driving and do
not depend on system dynamics, \textit{i.e.}, no back action of system
Floquet states on parameter modulations.
Therefore a natural and important question is whether novel Floquet physics
can emerge when system Floquet dynamics and parameter modulations are
mutually dependent.

In this Letter, we address this important question by studying Floquet
dynamics of ultracold boson atoms subject to a shaken optical lattice
generated inside an optical cavity. In the past several decades, the
interaction between atoms and static cavity fields with atom back actions
(no Floquet physics) have been well studied~in both theory~\cite%
{walther2006cavity} and experiment~\cite{PhysRevLett.91.203001,
PhysRevLett.98.053603, baumann2010dicke, PhysRevLett.107.140402,
PhysRevLett.109.153002, RevModPhys.85.553, PhysRevLett.113.070404,
PhysRevLett.114.123601, gopalakrishnan2009emergent, PhysRevLett.115.230403,
landig2016quantum}, showcasing rich cavity quantum electrodynamics (QED)
physics ranging from few-body problems such as Jaynes-Cummings model~\cite%
{shore1993jaynes} to many-body physics such as the Dicke superradiance~\cite%
{PhysRev.93.99, PhysRevLett.104.130401}. However, in these studies, the
cavity mode is static without periodic modulations such as shaking or moving.

Here we propose to realize a cavity-mode-induced freely evolving shaken
lattice, utilizing transverse pumping and a periodic modulation of the
cavity field phase, and study its mutual interaction with a non- or weakly
interacting Bose-Einstein condensate (BEC) inside the cavity.
While such shaken lattice generates Floquet bands for the BEC, the back
action of atom Floquet bands modulates the shaken lattice, leading to a
dynamical superradiant phase, where atom Floquet bands and shaken lattice
are self-adapted. Such Floquet normal-superradiant phase transition can be
dramatically different from non-Floquet one because of the coupling between
different Floquet sidebands. In particular, the interplay between intra- and
inter-sideband couplings may induce a hysteresis for the Floquet
normal-superradiant phase transition of non-interacting atoms, yielding a
completely new mechanism different from the well-known interaction-driven
hysteresis~\cite{lee2004first, gammelmark2011phase, luo2016dynamic,
Eckel2014, hamner2014dicke, trenkwalder2016quantum}. Surprisingly, the
steady state of the atom-cavity system can stabilize at the free energy
maximum for dominant inter-sideband coupling because of the non-equilibrium
nature of Floquet states. With increasing superradiant field, the Floquet
band dispersion gradually evolves from a single minimum to doubly degenerate
minima, leading to a second-order phase transition of the BEC.\ Such
transition can be significantly affected by the interaction between atoms
through the back action, which changes the critical superradiant field, the
Floquet band dispersion and the condensate momenta across the transition.

\begin{figure}[t]
\includegraphics[width=1.0\linewidth]{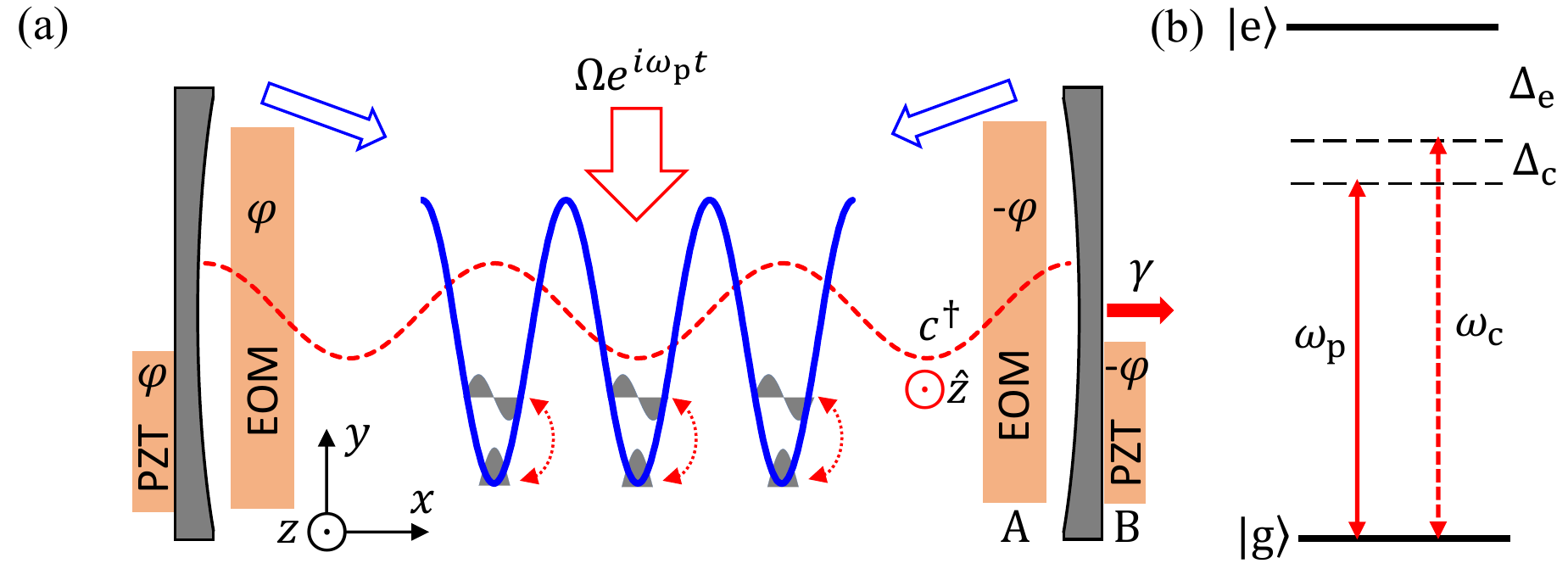}
\caption{ (a) Schematic of the experimental setup. The cavity is pumped by
an external transverse laser (red block arrow), with cavity mode (red line)
shaken by the two EOMs or PZT-driving mirrors. A BEC is prepared inside the
cavity with a background lattice (blue line) generated by additional lasers
(blue block arrows). The shaken cavity mode may induce inter-band couplings
(dotted arrows) when the band gap matches the shaking frequency. (b) Energy
levels of the atom and detunings of the cavity mode and pumping laser.}
\label{fig:sys}
\end{figure}

\emph{System.---}As shown in Fig.~\ref{fig:sys}(a), two schemes (A and B)
can be used to generate shaking cavity mode
as $\mathbf{\hat{z}}{E}(x,\varphi )\propto \mathbf{\hat{z}}\cos
[k_{0}x+\varphi (t)]$, with $\mathbf{\hat{z}}$ the polarization direction
and $k_{0}$ the wavenumber. Scheme A employs two electro-optic modulators
(EOMs)~\cite{yariv2007photonics}, while scheme B uses two mirrors
synchronously driven by piezoelectric transducers (PZTs)~\cite%
{goldovsky2016simple} to %are placed inside an optical cavity to
periodically and slowly [comparing with cavity free spectral range (FSR)]
modify the optical phase delay by $\varphi (t)=\varphi _{0}+f\cos (\omega
_{0}t)$.
The total optical path length does not change, therefore the cavity
resonance frequency is not affected. We consider a quasi-one-dimensional
(the dynamics in other directions are reduced by a deep harmonic trap) BEC
prepared in such a cavity, which is pumped by an external transverse laser.
The pumping frequency $\omega _{\text{p}}$ is close to the cavity resonance
frequency $\omega _{\text{c}}$, both of which are detuned far below the
atomic transition frequency $\omega _{\text{a}}$ [see Fig.~\ref{fig:sys}(b)].

\begin{figure}[t]
\includegraphics[width=1.0\linewidth]{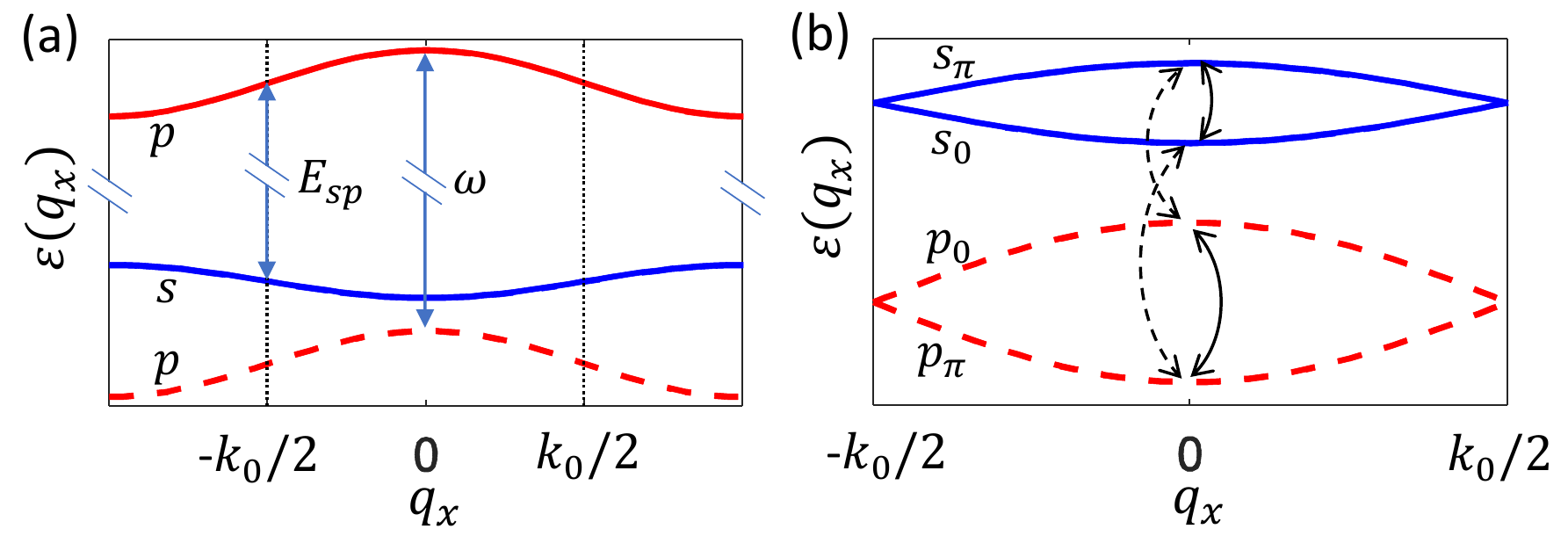}
\caption{(a) Illustration of atomic static bands (solid lines) and
phonon-dressed sidebands (dashed lines). (b) Couplings induced by $\hat{V}_{%
\text{c}}$ in Eq.~(\protect\ref{eq:vc}). The area of the Brillouin zone
reduces in half by $\hat{V}_{\text{c}}$. The time-independent
(time-dependent) term induces intra-sideband (inter-sideband) couplings
indicated by solid arrows (dashed arrows).}
\label{fig:band}
\end{figure}

After adiabatically eliminating the excited atomic level, we obtain the
Hamiltonian of the atom-cavity system in a rotating frame with $\hbar =1$
\begin{equation}
\mathcal{H}=(\Delta _{\text{c}}-u)c^{\dag }c+\int dx\Psi ^{\dag }(x)H_{\text{%
a}}(t)\Psi (x),  \label{Ham1}
\end{equation}%
where $c$ is the annihilation operator of the cavity photon, and $\Psi (x)$
is the matter wave field of atoms in the ground state. $\Delta _{\text{c}%
}=\omega _{\text{c}}-\omega _{\text{p}}$ is the cavity mode detuning, and $u=%
\frac{g_{0}^{2}}{\Delta _{\text{e}}}\int dx\Psi ^{\dag }\Psi \cos
^{2}(k_{0}x+\varphi )$ is the detuning induced by the atoms, which is
typically small and negligible. The single atom Hamiltonian is $H_{\text{a}%
}(t)=-\frac{\partial _{x}^{2}}{2m}+V_{\text{ext}}(x)+\hat{V}_{\text{c}}(x,t)$%
, with $V_{\text{ext}}(x)=v_{\text{e}}\cos ^{2}(k_{0}x)$ an static external
background lattice potential, which can be realized by additional lasers~%
\cite{SM}. It gives rise to a static tight-binding atomic band structure $%
\varepsilon _{\lambda }(q_{x})=E_{\lambda }+t_{\lambda }\cos (q_{x})$ with
band index $\lambda $ and Bloch momentum $q_{x}\in \lbrack -k_{0},k_{0}]$
[Fig. \ref{fig:band}(a)]. $\hat{V}_{\text{c}}(x,t)=-\eta \frac{c^{\dag }+c}{%
\sqrt{N_{\text{a}}}}\cos [k_{0}x+\varphi (t)]$ is the shaking potential
induced by the cavity-assisted ac-Stark shift, with $\eta =\Omega g_{0}\sqrt{%
N_{\text{a}}}/\Delta _{\text{e}}$ the coupling strength, $\Delta _{\text{e}%
}=\omega _{\text{a}}-\omega _{\text{p}}$ the single-photon detuning, $\Omega
$ and $g_{0}$ the Rabi frequencies of the transverse pumping laser and the
single cavity photon, respectively ($\Omega ,\text{ }g_{0}\ll \Delta _{\text{%
e}}$), and total atom number $N_{\text{a}}$.

Utilizing the expansion $e^{if\cos \omega
_{0}t}=\sum_{n}i^{n}J_{n}(f)e^{in\omega _{0}t}$ for $\cos
[k_{0}x+\varphi_0+f\cos(\omega_0t)]$, we see $\hat{V}_{\text{c}}(x,t)$ can
change the band structure by inducing a sequence of sidebands (\textit{i.e.}
phonon-dressed bands) and couplings between them (see Fig.~\ref{fig:band}).
We choose $v_{\text{e}}$, $\omega _{0}$, $\varphi _{0}$ and $f$ such that
the static $s$-band is near resonance with two-phonon-dressed ($\omega
\equiv 2\omega _{0}$) $p$-band [Fig.~\ref{fig:band}(a)], therefore only
these two bands need be considered for the calculation of new Floquet band
structure $\tilde{\varepsilon}_{\lambda }(q_{x})$. The first order expansion
$\propto e^{\pm i\omega _{0}t}$ is far-off resonance and can be neglected.
The zero-th order expansion corresponds to intra-sideband coupling, thus
only terms containing $\cos (k_{0}x)$ are nonzero due to the symmetry of the
Wannier functions. Similarly, only terms with $\sin (k_{0}x)$ are nonzero
for the second-order expansion $\propto e^{\pm i2\omega _{0}t}$ that couples
$s$ and $p$ bands. In total, the cavity-assisted potential can be written as~%
\cite{SM}%
\begin{equation}
\hat{V}_{\text{c}}=-\eta _{0}\frac{c^{\dag }+c}{\sqrt{N_{\text{a}}}}[\cos
(k_{0}x)+4\eta _{t}\cos (\omega t)\sin (k_{0}x)],  \label{eq:vc}
\end{equation}%
where $\eta _{0}=\eta J_{0}(f)\cos (\varphi _{0})$, $\eta _{t}=\frac{J_{2}(f)%
}{2J_{0}(f)}\tan (\varphi _{0})$ (tunable by $f$ and $\varphi _{0}$) is the
ratio between inter- and intra-sideband coupling strengths. Note that the
spatial period of $\hat{V}_{\text{c}}(x,t)$ is twice of $V_{\text{ext}}(x)$,
therefore the Brillouin zone (BZ) reduces by half to $q_{x}\in \lbrack
-k_{0}/2,k_{0}/2]$ through the band folding. Each band $\lambda $ is split
into two bands $\lambda _{0},\lambda _{\pi }$ [$\lambda =s,p$ as shown in
Fig.~\ref{fig:band}(b)] and the lattice potential $\hat{V}_{\text{c}}$ can
only couple 0 and $\pi $ bands due to the momentum transfer. $\varphi _{0}$
characterizes the relative phase between background lattice and the shaking
center of cavity field: for $\varphi _{0}=0$ ($\frac{\pi }{2}$), the shaking
potential is symmetric (asymmetric) at each site of the background lattice,
therefore can only induce intra-sideband (inter-sideband) couplings. Both
couplings coexist for $\varphi _{0}\neq j\pi /2$ ($j$ is an integer)~\cite%
{SM}.

\emph{Method.---}Under mean-field approximation,
the cavity field satisfies the Langevin equation $i\langle \dot{c}(t)\rangle
=\langle \left[ c,\mathcal{H}\right] \rangle -i\gamma /2\langle c(t)\rangle $
due to the weak leakage of the high-Q cavity,
yielding%
\begin{equation}
i\dot{\alpha}=\left( \Delta _{\text{c}}-u-i\gamma /2\right) \alpha -\eta
_{0}\Theta (t),  \label{eq:alpha}
\end{equation}%
with $\alpha (t)=\langle c(t)\rangle /\sqrt{N_{\text{a}}}$ and $\gamma $ the
cavity loss rate.
%due to $i\langle \dot{c}(t)\rangle =\langle \left[ c,\mathcal{H}\right] \rangle $,
Here $\Theta (t)=N_{\text{a}}^{-1}\int dx\langle \Psi ^{\dag }\Psi \rangle %
\left[ \cos (k_{0}x)+4\eta _{t}\sin (k_{0}x)\cos (\omega t)\right] $ is the
atomic density order. The frequency $\omega $ is chosen to be much larger
than $\eta _{0}\Theta $, $\Delta _{\text{c}}-u$, $\gamma /2$ (see section \emph{experimental consideration}), so that
high-order oscillation terms $\propto e^{\pm i\omega t}$ can be dropped in
Eq.~(\ref{eq:alpha}). The steady state solution in the presence of cavity
loss is determined by $\dot{\alpha}=0$, yielding%
\begin{equation}
\phi =2\tilde{\Delta}_{\text{c}}\eta _{0}\Theta _{0}/(\tilde{\Delta}_{\text{c%
}}^{2}+\gamma ^{2}/4),  \label{eq:alpha_s}
\end{equation}%
where $\phi \equiv \alpha +\alpha ^{\ast }$, $\tilde{\Delta}_{\text{c}%
}=\Delta _{\text{c}}-u_{0}$ is the effective detuning with $u_{0}=\frac{1}{T}%
\int_{0}^{T}dtu(t)$ a small constant, and $\Theta _{0}=\frac{1}{T}%
\int_{0}^{T}dt\Theta (t)$ with $T=2\pi /\omega $ the period. The shaking
cavity field $\phi $ and atom density order $\Theta _{0}$ are determined
self-consistently through Eq.~(\ref{eq:alpha_s}).

In the self-consistent determination, we replace $\frac{c^{\dag }+c}{\sqrt{%
N_{\text{a}}}}$ in $\hat{V}_{\text{c}}(x,t)$ [Eq.~(\ref{eq:vc})] by $\phi $
for the Floquet Hamiltonian $H_{\text{a}}(t)$ [Eq.~(\ref{Ham1})] and find
Floquet quasi-energy bands and Floquet states for the BEC, from which the
atom density order $\Theta $~can be calculated \cite{SM}. $\Theta $
in turn drives the cavity field $\phi $ through Eq.~(\ref{eq:alpha}) as the
atom feedback, yielding the self-adapted steady solution in Eq.~({\ref%
{eq:alpha_s}}). We find that solving the self-consistent equation Eq.~({\ref%
{eq:alpha_s}}) is equivalent to finding the extremum of the free energy
density $F(\phi )=\langle \mathcal{H}\rangle /N_{\text{a}}$ (\textit{i.e.}, $%
\frac{\partial F}{\partial \phi }=0$)~\cite{SM}. Notice that $\phi =\Theta
_{0}=0$ is always a trivial solution. Across the transition from normal to
superradiant phases, the zero solution becomes unstable and $\phi $, $\Theta
_{0}$ evolve from zero to finite values.

\begin{figure}[t]
\includegraphics[width=1.0\linewidth]{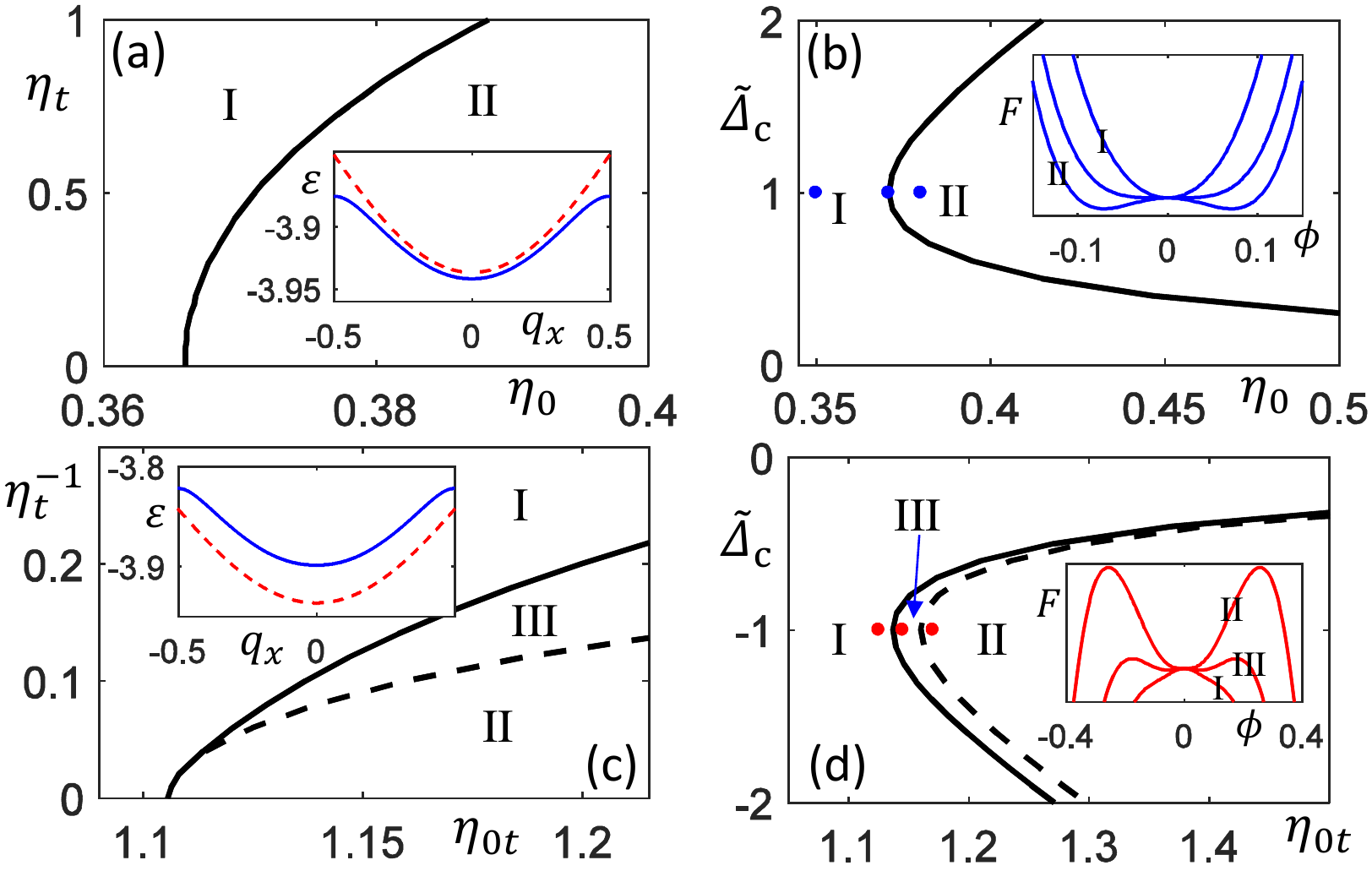}
\caption{(a) and (b) Phase diagrams for dominant intra-sideband coupling,
with $\tilde{\Delta}_{\text{c}}=1$ in (a) and $\protect\eta _{t}=0.5$ in
(b). I (II) represents the normal (superradiant) phase. (c) and (d) Phase
diagrams for dominant inter-sideband coupling, with $\tilde{\Delta}_{\text{c}%
}=-1$ in (c) and $\protect\eta _{t}^{-1}=0.1$ in (d), $\protect\eta _{0t}=%
\protect\eta _{0}\protect\eta _{t}$. III represents the hysteresis region.
The insets in (a) and (c) show the $s_{0}$-band dispersion in the normal
(red dashed line) and superradiant phase (blue solid line). The three blue
(red) dots in (b) [(d)] belong to different phases and they mark the points
where the free energy curves are plotted in the inset. The parameters are $%
v_{e}=-6$, $\protect\omega =4.8$ and $\protect\gamma =2$ (with energy unit $%
E_{\text{R}}=\hbar ^{2}k_{0}^{2}/2m$), and atom-atom interaction is
neglected.}
\label{fig:phase}
\end{figure}

\emph{Results.---}We focus on non-interacting BECs and discuss the
interaction effects later. The numerically calculated phase diagrams and
corresponding self-adapted Floquet bands with $\omega \gtrsim
E_{sp}+|t_{s}|+|t_{p}|$ are shown in Fig.~\ref{fig:phase}. For a small $\eta
_{t}$ ($\lesssim 1$), $\hat{V}_{\text{c}}$ is dominated by the
time-independent term that couples $s_{0}$ and $s_{\pi }$ bands, which would
lower the energy of $s_{0}$ band [see the inset in Fig.~\ref{fig:phase}(a)],
indicating $\langle \hat{V}_{\text{c}}\rangle =-\eta _{0}\phi \Theta _{0}<0$%
. According to Eq.~(\ref{eq:alpha_s}), a non-trivial steady state solution
exists only for blue effective detuning $\tilde{\Delta}_{\text{c}}>0$ [Figs.~%
\ref{fig:phase}(a) (b)]. The phase transition requires a stronger $\eta _{0}$
as $\eta _{t}$ increases, indicating that the transition becomes harder due
to the competition between inter- and intra-sideband couplings. As $\tilde{%
\Delta}_{c}$ decreases, the critical value of $\eta _{0}$ required for
superradiance first decreases then increases and tends to infinity at $|%
\tilde{\Delta}_{c}|=0$. This is because $\hat{V}_{\text{c}}$, which drives
the atomic density order $\Theta _{0}$, is proportional to $\phi $ and
approaches zero as $|\tilde{\Delta}_{c}|\rightarrow 0$ [see Eq.~(\ref%
{eq:alpha_s})]. We find that the solution is located at the minima of $%
F(\phi )$, which can be expanded as $F(\phi )=a_{2}\phi ^{2}+a_{4}\phi
^{4}+\cdots $. $F(\phi )$ exhibits a continuous transition from a single
minimum at $\phi =0$ to double minima at $\phi \neq 0$ [see the inset in
Fig.~\ref{fig:phase}(b)], where $a_{2}$ and $a_{4}$ are both positive before
the transition, and $a_{2}$ changes sign when the phase transition (second
order) occurs.

For a large $\eta _{t}$ ($\gg 1$), $\hat{V}_{\text{c}}$ is dominated by the
inter-band coupling between $s_{0}$ band and $p_{\pi }$ band that has a
lower energy than $s_{0}$ band, therefore atoms stays at the high-energy
excited band and increasing the cavity field would rise the band energy [the
inset in Fig.~\ref{fig:phase}(c)] of the BEC, leading to $\langle \hat{V}_{%
\text{c}}\rangle =-\eta _{0}\phi \Theta _{0}>0$. As a result, the
non-trivial steady state solution exists only for red effective detuning $%
\tilde{\Delta}_{\text{c}}<0$ [Figs.~\ref{fig:phase}(c) (d)]. Surprisingly,
the steady state solution is found at the maxima of $F(\phi )$ which
exhibits a transition from a single maximum at $\phi =0$ to double maxima at
$\phi \neq 0$ [the inset in Fig.~\ref{fig:phase}(d)] because of the
non-equilibrium phase transition of the dynamical steady states which may
not minimize the energy. Without Floquet sidebands, such superradiance at
energy maximum for $\tilde{\Delta}_{\text{c}}<0$ would not exist because
atoms generally prefer staying in the lowest static band which can only
couple with higher static bands.

\begin{figure}[t]
\includegraphics[width=1.0\linewidth]{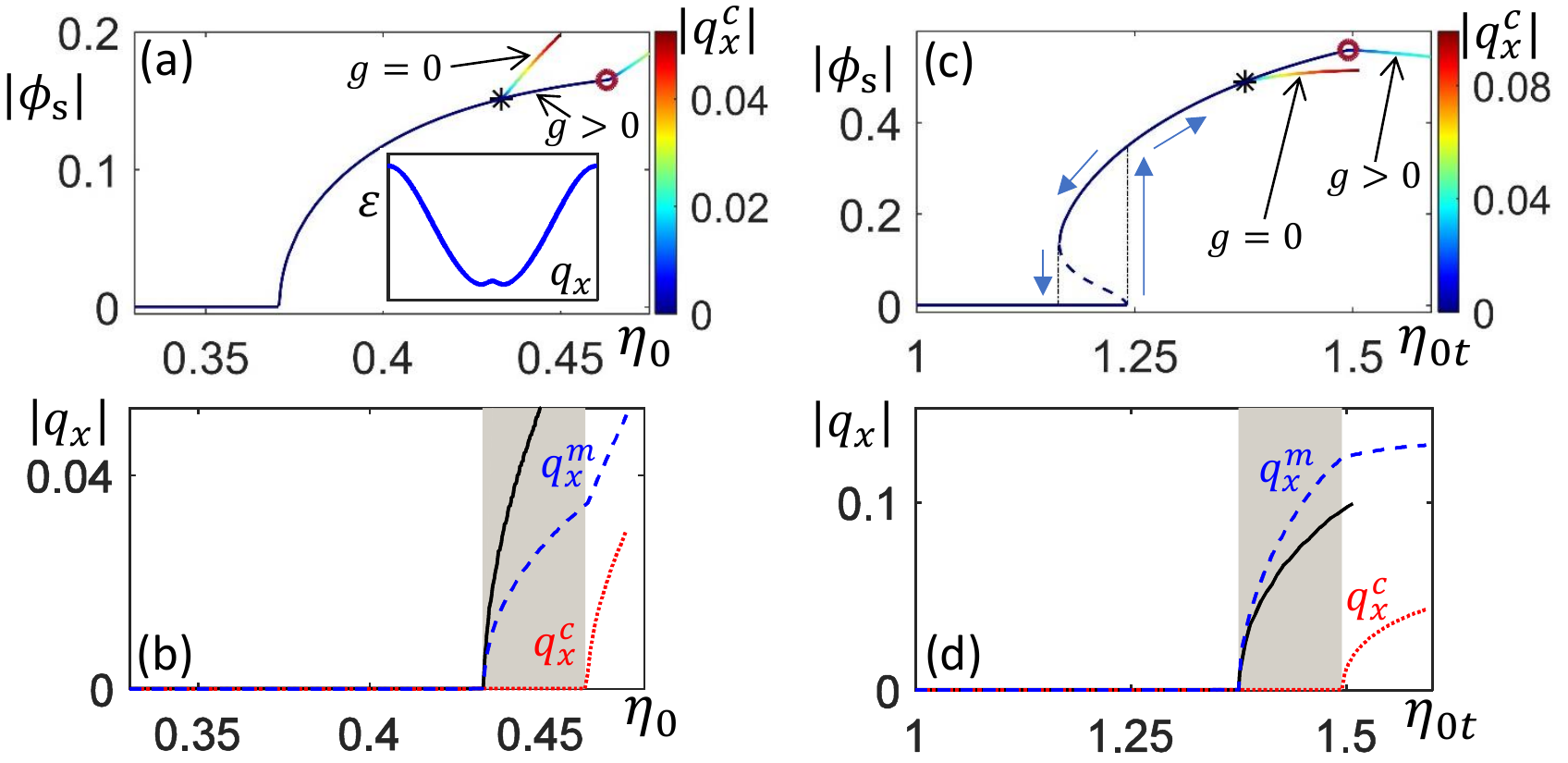}
\caption{(a) and (c) Superradiant order parameters and condensate momenta
versus $\protect\eta _{0}$ and $\protect\eta _{0t}$. Stars and circles mark
the transition points between zero and finite condensate momenta for $g=0$
and $g>0$, respectively. The inset of (a) shows the band structure for $%
|q_{x}^{m}|>0$. A clear hysteresis loop appears in (c) with solid lines
(dashed lines) corresponding to stable (unstable) steady-state solutions.
(b) and (d) Difference between $q_{x}^{c}$ (red dotted line) and $q_{x}^{m}$
(blue dashed line) for $g>0$. In the gray region, bosons are condensed into
the local band maximum at $q_{x}=0$. The black solid line shows $q_{x}^{m}=$
$q_{x}^{c}$ for $g=0$. (a) and (b) [(c) and (d)] correspond to dominant
intra-sideband (inter-sideband) coupling with $\Delta _{\text{c}}=1$, $%
\protect\eta _{t}=0.5$ and $g\bar{n}=0.02$ ($\Delta _{\text{c}}=-1$, $%
\protect\eta _{t}^{-1}=0.15$ and $g\bar{n}=0.1$). $\bar{n}=N_{\text{a}}/L$
is the average density with $L$ the system size. Other parameters are the
same as in Fig.~\protect\ref{fig:phase}. }
\label{fig:stable}
\end{figure}

Moreover, depending on the value of $\eta _{t}$, the transition can be
either continuous (second order) or discontinuous with a hysteresis loop
(first order) [Fig.~\ref{fig:stable}(b)]. Such hysteresis originates from
the interplay between intra-sideband and inter-sideband couplings, which
induces a second-order coupling (similar to a two-photon Raman process)
between $s_{0}$ and $p_{0}$ bands mediated by $s_{\pi }$ and $p_{\pi }$
bands (see Fig.~\ref{fig:band}(b)). Notice that the $p_{0}$ band is just
below the $s_{0}$ band, therefore this second-order coupling rises the $%
s_{0} $ band by increasing $a_{4}$. As a result, $a_{4}$ may change sign
(from negative to positive) prior to $a_{2}$ changes sign (from negative to
positive) when the second-order coupling is strong enough, leading to a
multi-stability behavior where $F(\phi )$ exhibits three maxima
simultaneously.

Hysteresis phenomena are related to strong nonlinearities~\cite%
{drazin1992nonlinear}, which are usually induced by strong atom-atom
interactions~\cite{lee2004first, Eckel2014}. For example, the strong Ising
interaction in the Dicke model can lead to a hysteresis loop of the
superradiant phase transition~\cite{lee2004first, gammelmark2011phase,
luo2016dynamic}. However, the hysteresis effect in our system does not need
atom-atom interaction at all, and has a completely different mechanism
originating from the coupling between Floquet sidebands induced by the
shaken cavity mode. Our study offers an excellent example and a realistic
system for observing hysteresis effects without atom interactions.

As the superradiant field increases, the $s_{0}$-$p_{0}$ band coupling would
induce a transition of the Floquet band $\tilde{\varepsilon}_{s_{0}}(q_{x})$
from a single minimum at $q_{x}=0$ to doubly degenerate minima at $q_{x}\neq
0$ [see inset in Fig.~\ref{fig:stable}(a)]. In our system, the $s_{0}$-$p_{0}
$ band coupling is a Raman-like coupling mediated by $s_{\pi }$ and $p_{\pi }
$, therefore such transition should be observed when the inter- and
intra-sideband couplings coexist. We consider atom-atom interaction (tunable
through Feshbach resonance~\cite{RevModPhys.82.1225}) that is weak and
repulsive, therefore the interaction energy is always minimized when the
system stays in a single momentum state. We assume that $\tilde{\varepsilon}%
_{s_{0}}(q_{x})$ is minimized at $\pm q_{x}^{m}$. For negligibly weak
interaction, the condensate momentum $q_{x}^{c}$ would locate at one of the
band minima (either $+q_{x}^{m}$ or $-q_{x}^{m}$) due to spontaneous
symmetry breaking~\cite{zheng2014strong}. Such a transition in the BEC would
also lead to a second order transition [stars in Figs.~\ref{fig:stable}(a)
(c)] of the shaking cavity field $\phi _{\text{s}}$ due to the back action.

As the atom-atom interaction increases (still weak enough such that the
superradiant phase transition is not affected and the long-time behaviors
(heating, atom loss, etc.)~are not significant \cite{parker2013direct,
khamehchi2016spin, PhysRevA.91.033601}), the phase
transition of the BECs from $q_{x}^{c}=0$ to $q_{x}^{c}\neq 0$ may be
shifted because $q_{x}^{c}\neq $ $\pm q_{x}^{m}$.  $q_{x}^{c}$ should be
determined by minimizing $\tilde{\varepsilon}_{s_{0}}(q_{x})+\tilde{%
\varepsilon}_{\text{int}}(q_{x})$, with $\tilde{\varepsilon}_{\text{int}%
}(q_{x})$ the momentum-dependent interaction energy,
%~\cite{zheng2014strong},
\begin{equation}
\tilde{\varepsilon}_{\text{int}}(q_{x})=\frac{1}{T}\int dt\int dx\frac{g}{2}%
|n_{s_{0},q_{x}}(x,t)|^{2}.
\end{equation}%
Here $g$ is the interaction constant and $n_{s_{0},q_{x}}(x,t)=\langle \Psi
^{\dag }\Psi \rangle $ is the atom density. In our system, band mixing
enhances spatial modulation of the density, and the interaction energy $%
\tilde{\varepsilon}_{\text{int}}(q_{x})$ is minimized at $q_{x}=0$ where the
mixing is the smallest. As a result, $q_{x}^{c}$ is smaller than $q_{x}^{m}$%
, and bosons may be condensed into the local maximum of the single-particle
band at $q_{x}=0$ [see Figs.~\ref{fig:stable}(b) (d)]. The transition from $%
q_{x}^{c}=0$ to $q_{x}^{c}\neq 0$ for $g>0$ requires a stronger
super-radiant field than the transition for $g=0$ [see Figs.~\ref{fig:stable}%
(a) (c)], and it also leads to a second order transition of $\phi _{\text{s}}
$ which in turn leads to a transition in $q_{x}^{m}$.

\emph{Experimental consideration.---}We consider a high-finesse (low-loss)
cavity with $\gamma =2E_{\text{R}}$ (with $E_{\text{R}}=\hbar
^{2}k_{0}^{2}/2m$ the recoil energy). Generally, atoms with a small mass are
preferred to obtain a large $E_{\text{R}}$, thus a large $\gamma $, which
makes the cavity easy to realize. For example, $^{7}$Li ($^{23}$Na) atoms
has a recoil energy $E_{\text{R}}\sim 40$kHz (10kHz), corresponding to $%
\gamma =80$kHz ($20$kHz), which can be realized with current technique~\cite%
{PhysRevLett.91.153003, SM}.
The shaking frequency $\omega_0$ is about several ten kHz for $^{23}$Na and several
hundred kHz for $^{7}$Li (both are much smaller than the free spectral range
$\sim $GHz), and such phase modulation can be implemented by PZT-driving
mirrors or low-loss
EOMs~\cite{SM}.
The system studied here only involves a change of introducing PZT-driving
mirrors or inserting two EOMs into the setups used in the ETH and Hamburg
laboratories~\cite{baumann2010dicke, PhysRevLett.113.070404,
PhysRevLett.115.230403, landig2016quantum, PhysRevLett.91.153003}, and thus
should be feasible with current technology. Moreover, our model can also be
implemented by combining a shaking external lattice and a periodic driving
force (e.g., using a periodically modulated magnetic field gradient)~\cite%
{SM}.

\emph{Discussion.---}We proposed a new type of Floquet physics where the
parameter modulations are not only related to external driving, but also
mutually coupled with system dynamics. As an example, we studied such
Floquet dynamics of BECs in shaking optical lattices, which lead to
interesting new phenomena including self-adapted shaking fields, Floquet
bands and hysteresis effects. Such self-adapted Floquet physics may also
arise in various other systems such as Rydberg-atom or molecule micromaser
\cite{walther1992experiments}, ion-trap cavity \cite{keller2004continuous}
and circuit quantum electrodynamics (circuit QED) \cite{wallraff2004strong},
etc. For example, in a circuit QED system, the electromagnetic field on the
waveguide resonator can be periodically modulated by attaching
superconducting quantum interference devices (SQUIDs) to the ends of the
resonator, with the SQUIDs driving by external magnetic fluxes~\cite%
{wilson2011observation}. Such a modulated waveguide resonator can couple
with superconducting qubits (artificial atoms) and these qubits may also
strongly couple with each other, where interesting self-adapted Floquet
dynamics may emerge. Such self-adapted Floquet circuit QED may find
important applications in quantum information processing and will be
addressed in future works. Within the shaking lattice cavity system,
interesting physics may also arise by considering strong atom-atom
interaction (Bose-Hubbard model~\cite{PhysRevLett.115.230403,
landig2016quantum}) or strong atom-single-photon coupling (limit cycles and
chaos~\cite{PhysRevLett.115.163601}), and superradiance of fermion gases
(topological bands~\cite{chen2014superradiance, pan2015topological}).
In this spirit, our proposal opens up new possibilities for studying
self-adapted Floquet physics in various systems, which may pave a way for
engineering new exotic quantum matter.

\begin{acknowledgments}
\textbf{Acknowledgements}: We thank A. Hemmerich for helpful discussion.
This work is supported by AFOSR (FA9550-16-1-0387), NSF (PHY-1505496), and
ARO (W911NF-17-1-0128).
\end{acknowledgments}

%\bibliographystyle{apsrev}
%\bibliographystyle{unsrt}
%\bibliography{Shaken_cavity_lattice}

\begin{widetext}
\section*{Supplementary Materials}
\setcounter{figure}{0} \renewcommand{\thefigure}{S\arabic{figure}} %
\setcounter{equation}{0} \renewcommand{\theequation}{S\arabic{equation}}

\subsection{Cavity-assisted shaking lattices and background lattices}

In this section, we give the detailed derivation of the cavity-assisted
shaking potential of Eq.~(2) in the main text, and discuss the possible
realizations of the background lattice.

The shaking cavity mode can be expanded using the Bessel function expansion $%
e^{if\cos \omega_0 t}=\sum_{n}i^{n}J_{n}(f)e^{in\omega_0 t}$,
\begin{eqnarray}
\cos [k_{0}x+\varphi _{0}+f\cos (\omega_0 t)] &=&\cos (k_{0}x)[J_{0}(f)\cos
(\varphi _{0})+2J_{1}(f)\sin (\varphi _{0})\cos (\omega_0 t)+2\cos (\varphi
_{0})J_{2}(f)\cos (2\omega_0 t)]  \nonumber \\
&&+\sin (k_{0}x)[J_{0}(f)\sin (\varphi _{0})+2J_{1}(f)\cos (\varphi
_{0})\cos (\omega_0 t)+2\sin (\varphi _{0})J_{2}(f)\cos (2\omega_0 t)]
\nonumber \\
&&+\ldots .
\end{eqnarray}%
We consider that only the static $s$ band and two-phonon-dressed $p$ band
are near resonance ($2\omega_0 \sim E_{sp}$), and the static lattice
potential $V_{\text{ext}}$ is much stronger than the shaking cavity field.
Due to the parity of the Wannier functions, $\sin (k_{0}x)$ only induces
inter-band couplings between $s$ and $p$ bands, while $\cos (k_{0}x)$ only
induces intra-band couplings.
%If the static $s$ band and one-phonon-dressed $p$ band are
%near resonance ($\omega_0 \sim E_{sp}$),
%and
With proper choice of the parameters $f$, $\varphi_{0} $ and $\omega_0$, the
expansion can be simplified by omitting small and far-off-resonance terms
%\begin{equation}
%\cos [k_{0}x+\varphi _{0}+f\cos (\omega_0 t)]\simeq J_{0}(f)\cos (\varphi
%_{0})[\cos (k_{0}x)+4\eta'_{t}\sin (k_{0}x)\cos (\omega_0 t)],
%\end{equation}
%with $\eta' _{t}=\frac{J_{1}(f)}{2J_{0}(f)}$, which can be tuned through $f$.
%A large $\eta'_{t}$ may require a large shaking amplitude $f$, which would
%enhance the high order terms in the Bessel function expansion and induce
%unwanted couplings.
%
%To overcome this problem, we may consider that the static $s$ band and the
%two-phonon-dressed $p$ band are near resonance ($\omega \equiv
%2\omega_0\sim E_{sp}$). Due to the parity of the Wannier functions,
%the expansion now can be simplified as
\begin{equation}
\cos [k_{0}x+\varphi _{0}+f\cos (\omega_0 t)]\simeq J_{0}(f)\cos (\varphi
_{0})[\cos (k_{0}x)+4\eta _{t}\sin (k_{0}x)\cos (\omega t)],
\end{equation}%
with $\eta _{t}=\frac{J_{2}(f)}{2J_{0}(f)}\tan (\varphi _{0})$ and $%
\omega\equiv2\omega_0$. Therefore, we obtain Eq.~(2) in the main text
\begin{equation}
\hat{V}_{\text{c}}=-\eta _{0}\frac{c^{\dag }+c}{\sqrt{N_{\text{a}}}}[\cos
(k_{0}x)+4\eta _{t}\cos \left( \omega t\right) \sin (k_{0}x)].
\end{equation}

The ratio $\eta _{t}$ between inter-sideband and intra-sideband couplings
can be tuned through $f$ and $\varphi _{0}$. For $\varphi _{0}=0$, we have $%
\eta _{0}=\eta J_{0}(f)$ and $\eta _{0}\eta _{t}=0$, corresponding to
dominant intra-sideband coupling. While for $\varphi _{0}=\frac{\pi }{2}$,
we have $\eta _{0}=0$ and $\eta _{0}\eta _{t}=\eta \frac{J_{2}(f)}{2}$,
corresponding to dominant inter-sideband coupling. These properties can be
understood from the parity of the cavity field with respect to the
background lattice sites. Fig.~\ref{figS:varphi}(a) shows the cavity field
for $\varphi _{0}=0$, where the static part of the cavity field is even with
respect to the background lattice site, thus induces strong intra-sideband
couplings due to the parity of the Wannier functions. While Fig.~\ref%
{figS:varphi}(b) shows the cavity field for $\varphi _{0}=\frac{\pi }{2}$,
where the static part of the cavity field is odd with respect to the
background lattice site, and cannot induce intra-sideband couplings. With
similar arguments, the shaking part, thus the inter-sideband coupling, is
dominant for $\varphi _{0}=\frac{\pi }{2}$.

\begin{figure}[tbp]
\includegraphics[width=0.6\linewidth]{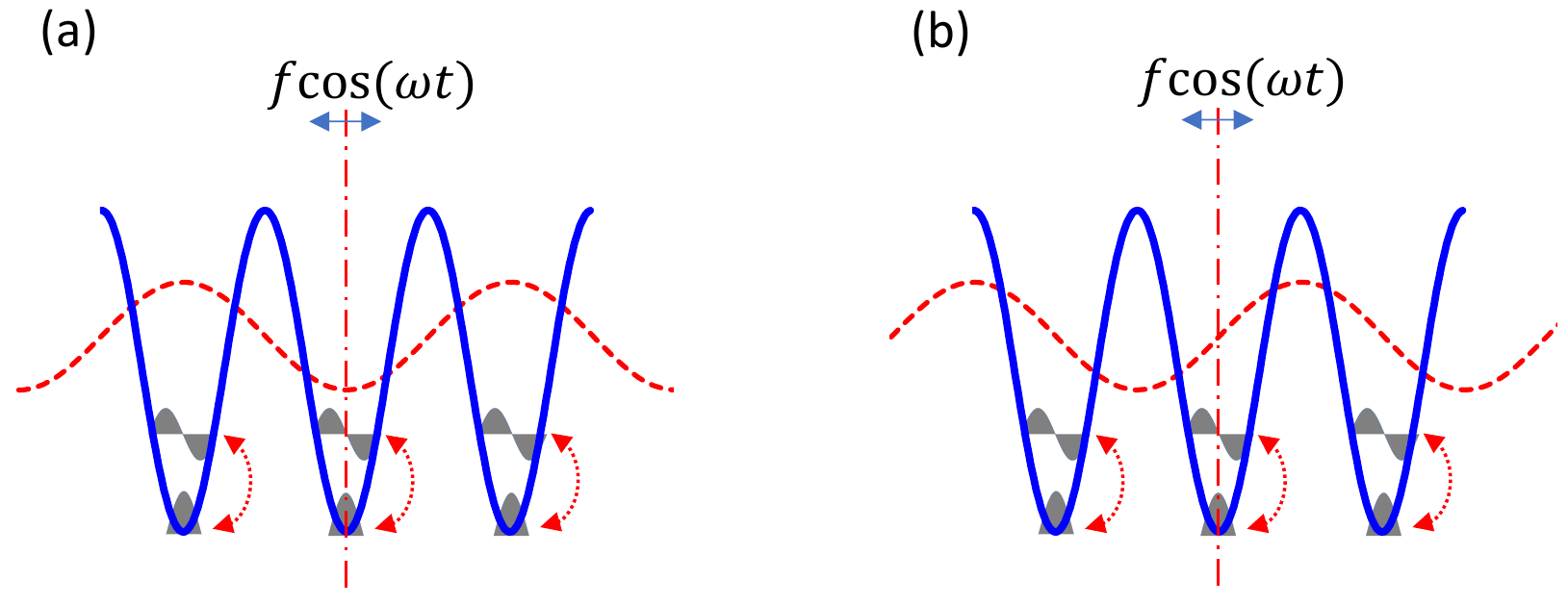}
\caption{ Parity of the cavity field (red dashed curve) with respect to the
back ground lattice (blue solid curve) for $\protect\varphi_0=0$ in (a) and $%
\protect\varphi_0=\frac{\protect\pi}{2}$ in (b).}
\label{figS:varphi}
\end{figure}

The above discussion can be generalized to the case with resonance coupling
between different side-bands. For example, we may consider the resonance
coupling between $s$ band and one-phonon-dressed $d$ band $\omega_0 \sim
E_{sd}$. Note that the coupling between $s$ and $d$ band is induced by $\cos
(k_{0}x)$, thus the shaken cavity mode can be simplified as
\begin{equation}
\cos [k_{0}x+\varphi _{0}+f\cos (\omega_0 t)]\simeq J_{0}(f)\cos (\varphi
_{0})\cos (k_{0}x)[1+4\eta^{\prime }_{t}\cos (\omega_0 t)],
\end{equation}
with $\eta^{\prime }_{t}=\frac{J_{1}(f)}{2J_{0}(f)}\tan (\varphi _{0})$,
which can be tuned through $f$ and $\varphi _{0}$.

The background lattice potential $V_{\text{ext}}(x)$ can be realized by
additional lasers feeding into the cavity from the transverse side with a
tilt angle $\theta $ and wavenumber $k_{0}/\cos (\theta )$~\cite%
{PhysRevLett.115.230403SS, landig2016quantumSS}. Alternatively, $V_{\text{ext}%
}(x)$ can be realized by longitudinally pumping the $\hat{\mathbf{y}}$%
-polarized cavity mode whose resonance frequency is far detuned from the $%
\hat{\mathbf{z}}$-polarized mode (so we may treat them independently)~\cite%
{gehr2010cavitySS, pan2015topologicalSS, landig2016quantumSS} (in this case,
one should use the EOMs to shake the cavity mode and the EOMs should only
modify the optical path length of the $\hat{\mathbf{z}}$-polarized mode~\cite%
{yariv2007photonicsSS}).

\subsection{Floquet bands and states for the atomic system}

Under mean-field approach, we replace operator $\left( c^{\dag }+c\right) /%
\sqrt{N_{\text{a}}}$ in $\hat{V}_{\text{c}}(x,t)$ by its mean value $\phi $,
the Hamiltonian of the atomic system can be written as
\begin{equation}
\mathcal{H}_{\text{a}}(\phi ,t)=\int dx\Psi ^{\dag }(x)[-\frac{\partial
_{x}^{2}}{2m}+V_{\text{ext}}(x)+{V}_{\text{c}}]\Psi (x),  \label{eq:Ha_s}
\end{equation}%
with %$H_0=-\frac{\partial_x^2}{2m}+V_\text{ext}(x)$ and
${V}_{\text{c}}=-\eta _{0}\phi \lbrack \cos (k_{0}x)+4\eta _{t}\sin
(k_{0}x)\cos (\omega t)]$. This Hamiltonian contains no photon operators,
and is a typical Hamiltonian for ultracold atoms in shaking lattices which
can be solved following the standard procedure. \emph{In this section}, we
show how to obtain the Floquet eigenstates and bands of the atomic system
for a given $\phi $, from which we can calculate the atomic density order $%
\Theta _{0}$ and obtain the steady state solution self-consistently [see
Eq.~(4) in the main text].

We consider a deep external lattice $V_{\text{ext}}(x)=v_{\text{e}}\cos
^{2}(k_{0}x)$ and a weak cavity field $\phi $. The %$H_0$ part is
time independent part $-\frac{\partial _{x}^{2}}{2m}+V_{\text{ext}}(x)$
leads to a static band structure, and ${V}_{\text{c}}$ as a perturbation
drives the resonance coupling between $s$ and $p$ bands. Here we will focus
on the dominant effects of the Floquet coupling and project the system onto
the subspace spanned by the bands near resonance. For a deep background
lattice $V_{\text{ext}}(x)$, we adopt the tight-binding approach and expand
the wave function in the Wannier basis of the $s$ and $p$ bands $\Psi
(x,t)=\sum_{\lambda ,j}a_{\lambda ,j}(t)W_{\lambda ,j}(x)$ with site index $j
$ and band index $\lambda =s,p$. The Hamiltonian is
\begin{equation}
\mathcal{H}_{\text{a}}(t)=\sum_{\lambda ,j}\left[ E_{\lambda }a_{\lambda
,j}^{\dag }a_{\lambda ,j}+t_{\lambda }(a_{\lambda ,j+1}^{\dag }a_{\lambda
,j}+h.c.)\right] +\sum_{\lambda ,j;\lambda ^{\prime },j^{\prime }}\Omega
_{\lambda ,\lambda ^{\prime }}^{j,j^{\prime }}(t)a_{\lambda ^{\prime
},j^{\prime }}^{\dag }a_{\lambda ,j}
\end{equation}%
with $\Omega _{\lambda ,\lambda ^{\prime }}^{j,j^{\prime }}(t)=\int
dxW_{\lambda ^{\prime },j^{\prime }}^{\ast }(x){V}_{\text{c}}(x,t)W_{\lambda
,j}(x)$ that drives the resonance coupling between $s$ and $p$ bands. The
cavity field would hardly affect the tunnelings between neighbor sites for a
deep background lattice, so we have $\Omega _{\lambda ,\lambda ^{\prime
}}^{j,j^{\prime }}\sim 0$ for $j\neq j^{\prime }$.

In the rotating frame with unitary transformation $U=\exp (-i\sum_{j}\omega
a_{p,j}^{\dag }a_{p,j}t)$, the Hamiltonian can be written as
\begin{equation}
\mathcal{H}_{\text{a}}(t)=\sum_{j}\Delta _{p}a_{p,j}^{\dag
}a_{p,j}+\sum_{\lambda ,j}\left[ t_{\lambda }(a_{\lambda ,j+1}^{\dag
}a_{\lambda ,j}+h.c.)+\Omega _{\lambda }^{j}(t)a_{\lambda ,j}^{\dag
}a_{\lambda ,j}\right] +\sum_{j}\left[ e^{-i\omega t}\Omega
_{s,p}^{j}(t)a_{s,j}^{\dag }a_{p,j}+h.c.\right] +N_{\text{a}}E_{s},
\end{equation}%
where $\Omega _{\lambda }^{j}(t)=\Omega _{\lambda ,\lambda }^{j,j}(t)=\int
dxW_{\lambda ,j}^{\ast }(x){V}_{\text{c}}(x,t)W_{\lambda ,j}(x)=-\eta
_{0}\phi \int dxW_{\lambda ,j}^{\ast }(x)\cos (k_{0}x)W_{\lambda ,j}(x)$,
and $\Omega _{s,p}^{j}(t)=\Omega _{s,p}^{j,j}(t)=\int dxW_{s,j}^{\ast }(x){V}%
_{\text{c}}(x,t)W_{p,j}(x)=-4\eta _{0}\eta _{t}\phi \cos (\omega t)\int
dxW_{s,j}^{\ast }(x)\sin (k_{0}x)W_{p,j}(x)$. $\Delta
_{p}=E_{p}-E_{s}-\omega $, and $N_{\text{a}}E_{s}$ is a constant. For a weak
cavity field with $\Omega _{\lambda ,\lambda ^{\prime }}^{j,j^{\prime }}\ll
\omega $, we drop the far-off-resonance terms and obtain the effective
time-independent Hamiltonian as (to the leading order)
\begin{equation}
\bar{\mathcal{H}}_{\text{a}}=\frac{1}{T}\int dt\mathcal{H}_{\text{a}%
}(t)=\sum_{j}\Delta _{p}a_{p,j}^{\dag }a_{p,j}+\sum_{\lambda ,j}\left[
t_{\lambda }(a_{\lambda ,j+1}^{\dag }a_{\lambda ,j}+h.c.)+\Omega _{\lambda
}^{j}a_{\lambda ,j}^{\dag }a_{\lambda ,j}\right] +\sum_{j}\left[ \Omega
_{s,p}^{j}a_{s,j}^{\dag }a_{p,j}+h.c.\right]
\end{equation}%
with $\Omega _{\lambda }^{j}=-\eta _{0}\phi \int dxW_{\lambda ,j}^{\ast
}(x)\cos (k_{0}x)W_{\lambda ,j}(x)$ and $\Omega _{s,p}^{j}=-2\eta _{0}\eta
_{t}\phi \int dxW_{s,j}^{\ast }(x)\sin (k_{0}x)W_{p,j}(x)$.

Notice that $\cos (k_{0}x)$ and $\sin (k_{0}x)$ have a spatial period which
is twice of that for the background potential $V_{\text{ext}}(x)$, thus we
have $\Omega _{\lambda }^{j}=(-1)^{j}\Omega _{\lambda }^{j=0}$ and $\Omega
_{s,p}^{j}=(-1)^{j}\Omega _{s,p}^{j=0}$. In the Fourier space (i.e., the
Bloch basis), $a_{\lambda ,q_{x}}\propto \sum_{j}e^{iq_{x}jb}a_{\lambda ,j}$
with $q_{x}\in \lbrack -k_{0},k_{0}]$ and $b=\frac{\pi }{k_{0}}$ the lattice
constant, we have
\begin{equation}
\bar{\mathcal{H}}_{\text{a}}=\sum_{\lambda ,q_{x}}\left[ \bar{\varepsilon}%
_{\lambda }(q_{x})a_{\lambda ,q_{x}}^{\dag }a_{\lambda ,q_{x}}+\Omega
_{\lambda }(q_{x})a_{\lambda ,q_{x}}^{\dag }a_{\lambda ,q_{x}+k_{0}}\right]
+\sum_{q_{x}}\left[ \Omega _{s,p}(q_{x})a_{s,q_{x}}^{\dag
}a_{p,q_{x}+k_{0}}+\Omega _{p,s}(q_{x})a_{p,q_{x}}^{\dag }a_{s,q_{x}+k_{0}}%
\right] ,
\end{equation}%
with $\bar{\varepsilon}_{s}(q_{x})=t_{s}\cos (q_{x}b)$, $\bar{\varepsilon}%
_{p}(q_{x})=\Delta _{p}+t_{p}\cos (q_{x}b)$, $\Omega _{\lambda
}(q_{x})=\Omega _{\lambda }^{j=0}$ and $\Omega _{s,p}(q_{x})=\Omega
_{p,s}(q_{x})=\Omega _{s,p}^{j=0}$. Since each unit cell now contains two
sites, the Brillouin zone (BZ) reduces in half and becomes $q_{x}\in \lbrack
-k_{0}/2,k_{0}/2]$. %and also the two bands are split into four bands.
Each band $\lambda $ is split into two bands $\lambda _{0},\lambda _{\pi }$,
with $a_{\lambda _{0}}(q_{x})=a_{\lambda }(q_{x})$ and $a_{\lambda _{\pi
}}(q_{x})=a_{\lambda }(q_{x}+k_{0})$. The Hamiltonian can be written as
\begin{equation}
\bar{\mathcal{H}}_{\text{a}}=(a_{p_{\pi }}^{\dag },a_{p_{0}}^{\dag
},a_{s_{0}}^{\dag },a_{s_{\pi }}^{\dag })\bar{H}_{\text{a}}(a_{p_{\pi
}},a_{p_{0}},a_{s_{0}},a_{s_{\pi }})^{T},
\end{equation}%
with
\begin{equation}
\bar{H}_{\text{a}}=\left(
\begin{array}{cccc}
\bar{\varepsilon}_{p_{\pi }}(q_{x}) & \Omega _{p}(q_{x}) & \Omega
_{s,p}(q_{x}) & 0 \\
\Omega _{p}^{\ast }(q_{x}) & \bar{\varepsilon}_{p_{0}}(q_{x}) & 0 & \Omega
_{p,s}^{\ast }(q_{x}) \\
\Omega _{s,p}^{\ast }(q_{x}) & 0 & \bar{\varepsilon}_{s_{0}}(q_{x}) & \Omega
_{s}(q_{x}) \\
0 & \Omega _{p,s}(q_{x}) & \Omega _{s}(q_{x}) & \bar{\varepsilon}_{s_{\pi
}}(q_{x})%
\end{array}%
\right) .
\end{equation}

Based on the time-independent Hamiltonian, we can obtain the Floquet $s_{0}$
band (i.e. the third band of $\bar{\mathcal{H}}_{\text{a}}$) with $\bar{%
\mathcal{H}}_{\text{a}}|{s_{0}},q_{x}\rangle =\tilde{\varepsilon}%
_{s_{0}}(q_{x})|s_{0},q_{x}\rangle $ and corresponding condensate atomic
density $n_{{s_{0}},q_{x}}(x,t)=\langle \Psi ^{\dag }(x,t)\Psi (x,t)\rangle
=\langle {s_{0}},q_{x}|U\Psi ^{\dag }(x,t)\Psi (x,t)U^{\dag }|{s_{0}}%
,q_{x}\rangle $, which can be used to calculate the atomic density order $%
\Theta _{0}$. In our numerical simulation, all matrix elements in $\bar{H}_{%
\text{a}}$ are calculated directly in the Bloch basis, where the effects of
small long-range tunnelings (e.g., $\Omega _{\lambda ,\lambda ^{\prime
}}^{j,j^{\prime }}$ for $j\neq j^{\prime }$) are also included. We
emphasize here that our system is different from a system with shaking
external lattices (which can not freely evolve) and static cavity fields,
where the proposed self-adapted Floquet dynamics and the hysteresis
phenomena would not exist due to the absence of cavity-assisted Floquet
sideband couplings.

\begin{figure}[tbp]
\includegraphics[width=0.6\linewidth]{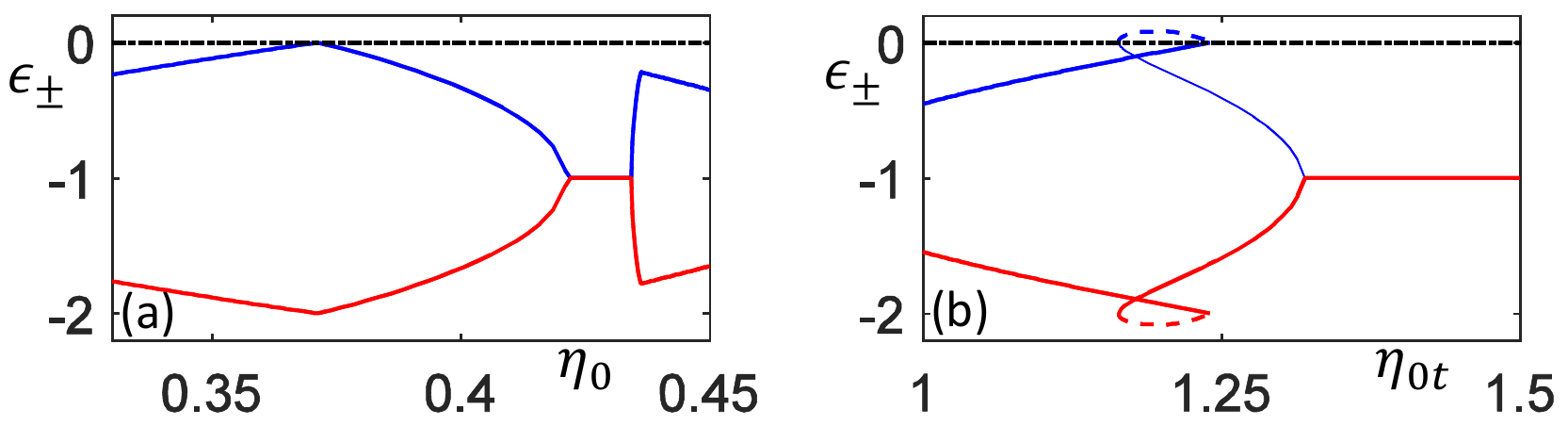}
\caption{Stability coefficients (blue and red solid lines) as a function of
the coupling strength for $g=0$, with black dash-dotted line representing
the zero axis. (a) corresponds to dominant intra-sideband coupling with $%
\Delta _{\text{c}}=1$ and $\protect\eta _{t}=0.5$. (b) corresponds to
dominant inter-sideband coupling with $\Delta _{\text{c}}=-1$ and $\protect%
\eta _{t}^{-1}=0.15$. A clear hysteresis loop appears in (b) with solid
lines (dashed lines) corresponding to stable (unstable) steady state
solutions. Other parameters are the same as in Fig.~4(c) in the main text. }
\label{fig:sm}
\end{figure}

\subsection{Energy density and self-consistent equation}

In this section, we show that solving the self-consistent equation is
equivalent to finding the extremum of the energy density.

The total Hamiltonian of the system is
\begin{equation}
\mathcal{H}=\tilde{\Delta}_{\text{c}}\ c^{\dag }c+\int dx\Psi ^{\dag }(x)[-%
\frac{\partial _{x}^{2}}{2m}+V_{\text{ext}}(x)+\hat{V}_{\text{c}}(x,t)]\Psi
(x),
\end{equation}%
with $\tilde{\Delta}_{\text{c}}=\Delta _{\text{c}}-u$. For a deep external
potential $V_{\text{ext}}(x)$, atoms are well localized near its minima,
therefore, the atom induced detuning becomes $u=\frac{g_{0}^{2}}{\Delta _{%
\text{e}}}\int dx\Psi ^{\dag }\Psi \cos ^{2}(k_{0}x+\varphi )\simeq \frac{%
g_{0}^{2}N_{\text{a}}}{\Delta _{\text{e}}}\cos ^{2}(\varphi )$, which is
small and independent of the cavity field. Under the mean-field approach,
the coupling term between atoms and the cavity field can be written as~\cite%
{luo2016dynamicSS}
\begin{eqnarray}
\int dx\Psi ^{\dag }(x)\hat{V}_{\text{c}}(x,t)\Psi (x) &=&-\eta _{0}\frac{%
c^{\dag }+c}{\sqrt{N_{\text{a}}}}\int dx\Psi ^{\dag }(x)\Psi (x)\left[ \cos
(k_{0}x)+2\eta _{t}(e^{i\omega t}+e^{-i\omega t})\sin (k_{0}x)\right]
\nonumber \\
&=&-\eta _{0}\frac{c^{\dag }+c}{\sqrt{N_{\text{a}}}}\int dx\langle \Psi
^{\dag }(x)\Psi (x)\rangle \left[ \cos (k_{0}x)+2\eta _{t}(e^{i\omega
t}+e^{-i\omega t})\sin (k_{0}x)\right]  \nonumber \\
&&-\eta _{0}\langle \frac{c^{\dag }+c}{\sqrt{N_{\text{a}}}}\rangle \int
dx\Psi ^{\dag }(x)\Psi (x)\left[ \cos (k_{0}x)+2\eta _{t}(e^{i\omega
t}+e^{-i\omega t})\sin (k_{0}x)\right]  \nonumber \\
&&+\eta _{0}\langle \frac{c^{\dag }+c}{\sqrt{N_{\text{a}}}}\rangle \int
dx\langle \Psi ^{\dag }(x)\Psi (x)\rangle \left[ \cos (k_{0}x)+2\eta
_{t}(e^{i\omega t}+e^{-i\omega t})\sin (k_{0}x)\right]  \nonumber \\
&=&-\eta _{0}\Theta (t)\sqrt{N_{\text{a}}}(c^{\dag }+c)+\eta _{0}\Theta (t){%
N_{\text{a}}}\phi -\int dx\Psi ^{\dag }(x)V_{\text{c}}\Psi (x).
\end{eqnarray}%
Thus the total Hamiltonian can be written as
\begin{equation}
\mathcal{H}=\tilde{\Delta}_{\text{c}}\ c^{\dag }c-\eta _{0}\Theta (t)\sqrt{%
N_{\text{a}}}(c^{\dag }+c)+\eta _{0}\Theta (t){N_{\text{a}}}\phi +\mathcal{H}%
_{\text{a}}(t).
\end{equation}%
with $\mathcal{H}_{\text{a}}(t)$ given by Eq.~(\ref{eq:Ha_s}).

We define the free energy $F$ as the total energy density
\begin{equation}
F=\frac{1}{N_{\text{a}}}\langle \mathcal{H}\rangle .
\end{equation}%
After dropping the far-off resonance terms, we obtain $F=\tilde{\Delta}_{%
\text{c}}\alpha ^{\ast }\alpha +\tilde{\varepsilon}_{s_{0}}(q_{x}^{c})$,
where we have used $\frac{1}{N_{\text{a}}T}\int dt\langle \mathcal{H}_{\text{%
a}}(t)\rangle \simeq \tilde{\varepsilon}_{s_{0}}(q_{x}^{c})$, with $\tilde{%
\varepsilon}_{s_{0}}(q_{x}^{c})$ the Floquet band obtained from the
time-independent Hamiltonian $\bar{\mathcal{H}}_{\text{a}}$. Using the
Langevin equation
\begin{equation}
\langle \dot{c}\rangle =-i\langle \lbrack c,\mathcal{H}]\rangle -\frac{%
\gamma }{2}\langle c\rangle =0,
\end{equation}%
we obtain %the self-consistent condition of Eq.~(4) in the main text, from
%which we have
$\tilde{\Delta}_{\text{c}}\alpha ^{\ast }\alpha =\frac{\tilde{\Delta}_{\text{%
c}}^{2}+\gamma ^{2}/4}{4\tilde{\Delta}_{\text{c}}}\phi ^{2}$. Then the
energy density can be written as
\begin{equation}
F=\frac{\tilde{\Delta}_{\text{c}}^{2}+\gamma ^{2}/4}{4\tilde{\Delta}_{\text{c%
}}}\phi ^{2}+\tilde{\varepsilon}_{s_{0}}(q_{x}^{c}).
\end{equation}%
Solving the self-consistent equation is equivalent to finding the extremum
of the free energy because
\begin{eqnarray}
\frac{\partial F}{\partial \phi } &=&\frac{\tilde{\Delta}_{\text{c}%
}^{2}+\gamma ^{2}/4}{2\tilde{\Delta}_{\text{c}}}\phi +\frac{\partial \tilde{%
\varepsilon}_{s_{0}}(q_{x}^{c})}{\partial \phi }  \nonumber \\
&=&\frac{\tilde{\Delta}_{\text{c}}^{2}+\gamma ^{2}/4}{2\tilde{\Delta}_{\text{%
c}}}\phi -\eta _{0}\Theta _{0}.
\end{eqnarray}%
Here we have used the Hellmann-Feynman theorem with $\frac{\partial \tilde{%
\varepsilon}_{s_{0}}(q_{x}^{c})}{\partial \phi }=\frac{-\eta _{0}}{N_{\text{a%
}}T}\int dtdx\langle \Psi ^{\dag }\Psi \rangle \lbrack \cos (k_{0}x)+4\eta
_{t}\sin (k_{0}x)\cos (\omega t)]=-\eta _{0}\Theta _{0}$. As a result, $%
\frac{\partial F}{\partial \phi }=0$ leads to
\begin{equation}
\phi =\frac{2\tilde{\Delta}_{\text{c}}}{\tilde{\Delta}_{\text{c}}^{2}+\gamma
^{2}/4}\eta _{0}\Theta _{0},
\end{equation}%
which is nothing but the self-consistent condition in Eq.~(4) in the main
text.

\subsection{Stability of the steady states}

The stability of the steady state solution can be determined from the
dynamical equation of the cavity field Eq.~(3) in the main text. Assuming a
small fluctuation $\delta \phi =\delta \alpha +\delta \alpha ^{\ast }$ in
the cavity field $\phi =\phi _{\text{s}}+\delta \phi $. Consequently, the
atomic density order obtains a fluctuation $\delta \Theta _{0}=\frac{%
\partial \Theta _{0}}{\partial \phi }\delta \phi $, yielding~\cite%
{bhaseen2012dynamicsSS, nagy2011criticalSS, luo2016dynamicSS}
\begin{equation}
\dot{\delta \alpha }=\left( -i\tilde{\Delta}_{\text{c}}-\frac{\gamma }{2}%
\right) \delta \alpha +i\eta _{0}\frac{\partial \Theta _{0}}{\partial \phi }%
(\delta \alpha +\delta \alpha ^{\ast }).  \label{eq:stable}
\end{equation}%
Eq.~(\ref{eq:stable}) has two solutions taking the form $\delta \alpha
=ae^{\epsilon _{\pm }t}+be^{\epsilon _{\pm }^{\ast }t}$, which are stable
(unstable) for $\Re (\epsilon _{\pm })<0$ [$\Re (\epsilon _{+})>0$ or $\Re
(\epsilon _{-})>0$]. The system would seek for the steady states that are
stable with exponentially decaying fluctuations %with decay coefficients
$\Re (\epsilon _{\pm })<0$, corresponding to the energy minima (maxima) for $%
\eta _{t}\ll 1$ and $\Delta _{\text{c}}>0$ ($\eta _{t}\gg 1$ and $\Delta _{%
\text{c}}<0$), as shown in Figs.~\ref{fig:sm}(c) (d). At the boundary of
superradiant phase transition (or the boundary of the hysteresis loop), one
of the decay coefficients tends to zero as expected. The transition from
zero to finite condensate momentum is characterized by the order parameter $%
q_{x}^{c}$, thus the decay coefficients of $\delta \phi $ near this
transition need not tend to zero.

\subsection{Experimental consideration}

\emph{Piezoelectric transducers scheme.---}The shaking cavity mode can be
realized by mounting two piezoelectric transducers (PZTs) on the two mirrors
of the cavity (one on each), and synchronously driving the piezoelectric
transducers (PZTs). Alternatively, one can externally drive one PZT and lock
the cavity using the other PZT. in this case, the cavity locking and shaking
are achieved simultaneously. The servo bandwidth of PZTs are typically of
the order of 10kHz, which can be improved significantly using specially
designed mounting. For instance, PZTs with servo bandwidth of 200kHz and
500kHz have been demonstrated~in experiments \cite{goldovsky2016simpleSS,
nakamura2017beyondSS}. As a result, a shaking frequency up to several hundred
kHz can be realized using such PZTs. We can use $^{7}$Li atoms and a
high-finesse cavity with a loss rate $\gamma \lesssim 100$kHz which can be
realized using high reflection mirrors~\cite{PhysRevLett.91.153003SS}.

\emph{Electro-optic optical modulator scheme.---}The shaking cavity mode can
also be realized by inserting two EOMs into the cavity which modify the
optical phase delay through tuning the applied voltage. We have considered a
high-finesse cavity with a loss rate $\gamma \lesssim 100$kHz (correspond to
a finesse of $10^{4}$ for a cavity with free spectral range $\sim 1$GHz).
While cavities without EOMs inside can reach a finesse of $10^{5}$ (with
round trip loss $<0.005\%$)~in experiments \cite{PhysRevLett.91.153003SS}.
%corresponding to $\gamma \sim 10$ kHz for cavities with free spectral range (FSR) $\sim $GHz.
It has been demonstrated that a highly transmissive intra-cavity element
(realized by anti-reflection coating or Brewster's angle incidence) can
hardly reduce the finesse of the cavity~\cite{PhysRevLett.94.193201SS,
warrier2014highlySS}.
% (corresponding to a loss rate of a few kHz for cavities with FSR of the order of GHz).
As a result, to maintain the high finesse of a cavity with EOMs inside, the
EOMs should have extremely low losses (high transmission and low
absorption). Using high-efficiency anti-reflection (AR) coating, the
reflectivity at target wavelength can be smaller than $0.01\%$~\cite%
{raut2011antiSS}, leading to a round-trip loss about $\sim 0.04\%$. Such low
reflectivity can also be realized by Brewster's angle incidence~\cite%
{PhysRevLett.94.193201SS} (the Brewster's angle would hardly affected by the
applied voltage on the EOM since the change of reflectivity index is less
than $10^{-4}$).
%For a cavity with FSR $\sim0.5$GHz, this would induce an extra cavity loss $\lesssim30$kHz.
The absorption of the EOM can be suppressed by using low-absorption
electro-optic materials (e.g., gray-track resistance potassium titanyl
phosphate crystal or magnesium-doped lithium niobate crystal with an
absorption coefficient $\sim 10^{-4}$cm$^{-1}$~\cite{roth2001opticalSS,
leidinger2015highlySS, waasem2013photoacousticSS, yariv1984opticalSS}). We also
notice that $\phi _{0}$ can be viewed as the phase delay without applying
voltage on the EOM, which is used to tune the relative position between the
cavity field and the external lattice potential and can also be achieved by
shifting the external lattice. Therefore the amplitude of the EOM phase
modulation is $f$ which is small $f\sim 0.1\pi $ (enough for our purpose),
and the length of the EOM can be reduced to $<1$cm using typical voltage
driving (with an amplitude of a few hundred Volts)~\cite{yariv1984opticalSS}.
The absorption loss per round trip is also of the order of $10^{-4}$, which
can be improved further by using higher voltage driving since the modulation
frequency is only of several hundred kHz).
% (corresponding to an extra cavity loss $\sim10$kHz),
As a result, the total round-trip loss of the cavity can be as low as $\sim
0.05\%$, corresponding to a finesse about 12000. Moreover, to reduce
requirement for the finesse, it is also possible to use a longer cavity with
smaller free spectral range (FSR), for example, a cavity with a FSR of
0.5GHz only needs a finesse of 5000 to achieve the required loss rate $100$%
kHz.

\emph{Gradient driving force scheme.---}Finally, we notice that by combing a
shaking external lattice $V_{\text{ext}}=v_{\text{e}}\cos ^{2}[k_{0}x-f\cos
(\omega _{0}t)]$ and a periodic driving force from a periodically modulated
magnetic field gradient $B=B^{\prime }x\cos (\omega _{0}t)$~\cite%
{PhysRevLett.104.200403SS}, it is possible to realize the same Hamiltonian as
we proposed. In the co-moving frame of the shaking external lattice, the
cavity lattice (static in the laboratory frame) is effectively shaking. Due
to the equivalence between acceleration and force~\cite%
{PhysRevLett.108.225304SS}, the periodic force is canceled out if
\begin{equation}
\frac{f}{k_{0}}m\omega _{0}^{2}=m_{F}g_{F}\mu _{B}B^{\prime },
\end{equation}%
where $m$ is atom mass, $m_{F}$ is the angular momentum quantum number along
the quantized axis, $g_{F}$ is Lande g-factor and $\mu _{B}$ the Bohr
magneton. With these parameters, the amplitude magnetic field gradient
should be of the order of $10^{6}$Gauss/m. Currently, periodically driving
magnetic field gradient with amplitude $10^{4}$Gauss/m and a frequency of
several kHz has been realized~\cite{luo2016tunableSS}. Though in principle
this could be improved using better designed (anti-)Helmholtz coils (with
more winding turns) and driving current sources (with larger amplitude and
frequency), a $10^{6}$Gauss/m gradient with modulation frequency at several
ten kHz might be still challenging.
%if their phases, amplitudes and frequencies are match, and

\emph{Summary.---}The EOM and PZT-driving mirror implementations discussed
above are feasible with current technique, while the PZT driving mirror
requires a slightly simpler experimental setup than the EOM one. The
combination of a shaking external lattice and a periodic driving force would
require a more complex setup than the other two and might be still
challenging to realize.

%For a cavity with FSR $\sim
%1$GHz, a total cavity loss rate $\gamma \lesssim 100$kHz thus can be
%achieved.

\end{widetext}

\end{document}